\documentclass{article}
\usepackage{amsmath, amssymb, amsfonts}
\title{ On a nonstatic Painleve-Gullstrand spacetime } 
\author{Hristu Culetu, \\Ovidius University, Department of Physics and Electronics, \\ Bld. Mamaia  124, 900527 Constanta, Romania, \\e-mail : hculetu@yahoo.com}

\begin{document}
\numberwithin{equation}{section}
\pagenumbering{arabic}
\maketitle
\newcommand{\fv}{\boldsymbol{f}}
\newcommand{\tv}{\boldsymbol{t}}
\newcommand{\gv}{\boldsymbol{g}}
\newcommand{\OV}{\boldsymbol{O}}
\newcommand{\wv}{\boldsymbol{w}}
\newcommand{\WV}{\boldsymbol{W}}
\newcommand{\NV}{\boldsymbol{N}}
\newcommand{\hv}{\boldsymbol{h}}
\newcommand{\yv}{\boldsymbol{y}}
\newcommand{\RE}{\textrm{Re}}
\newcommand{\IM}{\textrm{Im}}
\newcommand{\rot}{\textrm{rot}}
\newcommand{\dv}{\boldsymbol{d}}
\newcommand{\grad}{\textrm{grad}}
\newcommand{\Tr}{\textrm{Tr}}
\newcommand{\ua}{\uparrow}
\newcommand{\da}{\downarrow}
\newcommand{\ct}{\textrm{const}}
\newcommand{\xv}{\boldsymbol{x}}
\newcommand{\mv}{\boldsymbol{m}}
\newcommand{\rv}{\boldsymbol{r}}
\newcommand{\kv}{\boldsymbol{k}}
\newcommand{\VE}{\boldsymbol{V}}
\newcommand{\sv}{\boldsymbol{s}}
\newcommand{\RV}{\boldsymbol{R}}
\newcommand{\pv}{\boldsymbol{p}}
\newcommand{\PV}{\boldsymbol{P}}
\newcommand{\EV}{\boldsymbol{E}}
\newcommand{\DV}{\boldsymbol{D}}
\newcommand{\BV}{\boldsymbol{B}}
\newcommand{\HV}{\boldsymbol{H}}
\newcommand{\MV}{\boldsymbol{M}}
\newcommand{\be}{\begin{equation}}
\newcommand{\ee}{\end{equation}}
\newcommand{\ba}{\begin{eqnarray}}
\newcommand{\ea}{\end{eqnarray}}
\newcommand{\bq}{\begin{eqnarray*}}
\newcommand{\eq}{\end{eqnarray*}}
\newcommand{\pa}{\partial}
\newcommand{\f}{\frac}
\newcommand{\FV}{\boldsymbol{F}}
\newcommand{\ve}{\boldsymbol{v}}
\newcommand{\AV}{\boldsymbol{A}}
\newcommand{\jv}{\boldsymbol{j}}
\newcommand{\LV}{\boldsymbol{L}}
\newcommand{\SV}{\boldsymbol{S}}
\newcommand{\av}{\boldsymbol{a}}
\newcommand{\qv}{\boldsymbol{q}}
\newcommand{\QV}{\boldsymbol{Q}}
\newcommand{\ev}{\boldsymbol{e}}
\newcommand{\uv}{\boldsymbol{u}}
\newcommand{\KV}{\boldsymbol{K}}
\newcommand{\ro}{\boldsymbol{\rho}}
\newcommand{\si}{\boldsymbol{\sigma}}
\newcommand{\thv}{\boldsymbol{\theta}}
\newcommand{\bv}{\boldsymbol{b}}
\newcommand{\JV}{\boldsymbol{J}}
\newcommand{\nv}{\boldsymbol{n}}
\newcommand{\lv}{\boldsymbol{l}}
\newcommand{\om}{\boldsymbol{\omega}}
\newcommand{\Om}{\boldsymbol{\Omega}}
\newcommand{\Piv}{\boldsymbol{\Pi}}
\newcommand{\UV}{\boldsymbol{U}}
\newcommand{\iv}{\boldsymbol{i}}
\newcommand{\nuv}{\boldsymbol{\nu}}
\newcommand{\muv}{\boldsymbol{\mu}}
\newcommand{\lm}{\boldsymbol{\lambda}}
\newcommand{\Lm}{\boldsymbol{\Lambda}}
\newcommand{\opsi}{\overline{\psi}}
\renewcommand{\tan}{\textrm{tg}}
\renewcommand{\cot}{\textrm{ctg}}
\renewcommand{\sinh}{\textrm{sh}}
\renewcommand{\cosh}{\textrm{ch}}
\renewcommand{\tanh}{\textrm{th}}
\renewcommand{\coth}{\textrm{cth}}

\begin{abstract}
A time dependent geometry outside a spherically symmetric mass is proposed. The source has zero energy density but nonzero radial and tangential pressures. The time variable is interpreted as the duration of measurement performed upon the physical system. For very short time intervals, the effect of the mass source is much reduced, going to zero when $t \rightarrow 0$. All physical quantities are finite when $t \rightarrow 0$ and $r \rightarrow 0$ and also at infinity. The total energy flux measured on a hypersurface of constant $r$ is vanishing.\\
\textbf{Keywords}: variable mass; anisotropic source; energy flux, decoherence time 
 \end{abstract}
 
 \section{Introduction}
 It is a known fact that one of the conceptual problems of quantum theory is the so-called ''measurement problem'' - the standard Quantum Mechanics (QM) crucially depends on the concept of measurement, even though such notion is not defined rigurously within the theory \cite{OS, NG, GF}. According to Okon and Sudarsky, the solution to the measurement problem may well lie in Quantum Gravity (QG), which is still lacking. Moreover, they suggest that it may be necessary to solve the measurement problem for to build a quantum theory of gravity. Gisin \cite{NG} and Gisin and Frowis \cite{GF} argued that, without solving the measurement problem, quantum theory is not complete, as it does not tell us how one should - in principle - perform measurements. They consider the time is ready to pass from the study of Quantum non-locality - a very fruitful subject of research - to the Quantum measurement problem, another basic problem in the foundations of Quantum Mechanics.

 Connections between quantum-foundational issues and QG have been also pointed out by Penrose \cite{RP} (see also \cite{LD1, LD2, DB, GGB}), who studied the intrinsic spacetime instability when macroscopic bodies are placed in quantum superposition of different locations, an idea that leads him to a link between quantum collapse of the wave function and gravity. Diosi \cite{LD3} introduced a nonlocal, gravitational term in the time-dependent Schrodinger equation for to find the quantum uncertainty in the position of a free pointlike macroscopic object, from the minimization of the energy. In addition, Pearle and Squires \cite{PS} suggested that the curvature scalar of the spacetime is responsable for the spontaneous quantum collapse.

 Motivated by the importance of the measurement process within QM and QG, we investigate in this paper the role played by the duration of measurement on the spacetime structure of the physical system under consideration. We know that the (Newtonian) gravity has not been checked out experimentally in the range below $0.1$ mm. We pass from short range distances to short range time intervals and suggest that the strength of the gravitational field may be modified when the measurement is performed in a very short time interval w.r.t. the gravitational radius of the object.

 Throughout the paper we use geometrical units $G = c = \hbar = 1$, unless otherwise specified.

\section{Painleve-Gullstrand geometry with time dependent mass}
 The Schwarzschild exact solution for the geometry outside a star or a BH is given by
   \begin{equation}
  ds^{2} = -(1- \frac{2m}{r}) dt_{S}^{2} + (1- \frac{2m}{r})^{-1} dr^{2} + r^{2} d \Omega^{2}. 
 \label{2.1}
 \end{equation}
 In (2.1) $t_{S}$ is the Schwarzschild time and $d\Omega^{2}$ is the metric on the unit 2-sphere. To get rid of the coordinate singularity of the metric at the horizon $r = 2m$, Painleve and Gullstarnd (P-G) used the following temporal transformation \cite{KW, TWZ, EP}
   \begin{equation}
   t = t_{S} + 2\sqrt{2mr}+ 2m ln\frac{\sqrt{r} - \sqrt{2m}}{\sqrt{r} + \sqrt{2m}}. 
\label{2.2}
 \end{equation}
Therefore, the line element appears as 
   \begin{equation}
  ds^{2} = -(1- \frac{2m}{r}) dt^{2} + dr^{2} + 2\sqrt{\frac{2m}{r}} dt dr + r^{2} d \Omega^{2}. 
 \label{2.3}
 \end{equation}
 where $t$ is the free-fall time, that is the proper time experienced by an observer who free-falls from rest at infinity.
We chose the ''+'' sign in front of the square root in order to deal only with the inward moving free particles (along a geodesic curve with $dr + \sqrt{2m/r}dt = 0$, the velocity $dr/dt = - \sqrt{2m/r} $ is negative).

The geometry (2.3) is stationary, namely invariant under time translations (however, it is not invariant under time reversal because of the nondiagonal term). In addition, a constant time slice is simply flat space. We also emphasize that (2.3) represents physical space freely falling radially into the BH at the Newtonian escape velocity $\sqrt{2m/r}$. The proper time of one observer at rest ($dr = d \theta = d \phi = 0$) is $d \tau = \sqrt{1 - (2m/r)}dt$.

 As we know, a time dependent source with spherical symmetry will no longer lead to a Ricci-flat geometry, i.e., to a vacuum solution of the Einstein equations. Therefore, Birkhoff's theorem does not apply for this case. A nonstatic Schwarzschild (S) spacetime with a time dependent mass, outside an object with spherical symmetry was investigated in \cite{HC}. It was found there that the source of geometry (an anisotropic fluid) has zero energy density and radial pressure, nonzero tangential pressures and radial energy flux. We intend in this paper to introduce a variable mass directly in the line-element (2.3), for to explore whether more simple properties may be obtained. Therefore, we write down the geometry (2.3) as
   \begin{equation}
  ds^{2} = -(1- \frac{2m e^{-\frac{k}{t}}}{r}) dt^{2} + dr^{2} + 2\sqrt{\frac{2m e^{-\frac{k}{t}}}{r}} dt dr + r^{2} d \Omega^{2},
 \label{2.4}
 \end{equation}
with $lne = 1,~m(t) = me^{-\frac{k}{t}},~t>0$,  $k$ - positive constant and $m$ - the particle constant mass. To find $k$, we make use of reasonings from \cite{RP, LD1, LD2, LD3}: one looks for a link between quantum collapse of the wave function and gravity, when macroscopic objects are placed in quantum superposition at different locations. Diosi \cite{LD3} added a nonlocal gravitational term to the standard QM terms from the Schrodinger equation
   \begin{equation}
	i\hbar \frac{\partial \Psi(\textbf{x},t)}{\partial t} = - \frac{\hbar^{2}}{2M} \Delta \Psi - GM^{2} \int \frac{|\Psi(\textbf{x'},t)|^{2}}{|\textbf{x} - \textbf{x'}|} d^{3}\textbf{x'}~ \Psi(\textbf{x},t), 
 \label{2.5}
 \end{equation} 
for a macroscopic object of mass $M$ and radius $R$, with $\textbf{x}$ and $\textbf{x'}$ - the locations of the two branches of the superposition. Diosi showed that, when $\Delta \textbf{x}\equiv|\textbf{x} - \textbf{x'}| << R$, the Newtonian potential energy from (2.5) acquires the form
    \begin{equation}
		U(\textbf{x} - \textbf{x'}) \approx U(0) + \frac{1}{2} M\omega^{2} |\textbf{x} - \textbf{x'}|^{2},
 \label{2.6}
 \end{equation} 
where $\omega^{2} = GM/R^{3} = 4\pi G\rho/3$ is the frequency of the Newtonian oscillator (which could be obtained from the geodesic deviation), $\rho$ is the constant density of the particle and $U(0) = GM^{2}/R$. The standard kinetic term from (2.5) tends to spread the wave function, competing with the Diosi-Penrose spontaneous collapse which tends to shrink the wave function. When the spreading rate $\hbar/M(\Delta \textbf{x})^{2}$ equals the collapse rate $1/\tau \equiv M\omega^{2} (\Delta \textbf{x})^{2}/\hbar$, an equilibrium is reached and one obtains $1/\tau = \omega$, where $\tau$ represents the decoherence time, required to collapse the macroscopic superposition, or the quantum Zenon time \cite{LD4}. 

 For our case of interest, we propose to consider $\tau$ as the time that light needs to cross the Schwarzschild radius of the object. In this case we have $\tau = 2M$ which means to insert $k = 2m$ in Eq. (2.4). Hence (2.4) becomes
   \begin{equation}
  ds^{2} = -(1- \frac{2m e^{-\frac{2m}{t}}}{r}) dt^{2} + dr^{2} + 2\sqrt{\frac{2m e^{-\frac{2m}{t}}}{r}} dt dr + r^{2} d \Omega^{2},
 \label{2.7}
 \end{equation}
with $m(t) = m e^{-\frac{2m}{t}}$. To avoid a signature switch of the metric coefficient $g_{tt}$, we impose the condition $f(r,t) \equiv 1- \frac{2m}{r} e^{-\frac{2m}{t}} >0$, namely $r > 2m e^{-\frac{2m}{t}}$, with $r_{AH} = 2m e^{-\frac{2m}{t}}$ - the location of the apparent horizon. That is necessary because otherwise the proper time and $t$ will not have the same sign for an observer located somewhere at $r, \theta, \phi = const$. For constant $r$, $f(r,t)$ is a monotonic decreasing function of $t$, tends to unity when $t \rightarrow 0$ and acquires the standard Schwarzschild value $(1 - 2m/r)$ at infinity (or when $t >> 2m$). When $f(r,t)$ is considered as a function of $r$, it equals unity for $r \rightarrow \infty$. However, the limit $r \rightarrow 0$ has to be taken with $t \rightarrow 0$, in order to satisfy the condition $r > 2m(t)$. Consequently, $0 < f(r,t) <1$. We notice also that the apparent horizon is an increasing function of $t$, from $r_{AH} \rightarrow 0$ when $t \rightarrow 0$ and $r_{AH} \rightarrow 2m$ at infinity, having an inflexion point at $t = m$.

We take the timelike variable $t$ as the duration of measurement, so that from (2.7) results that gravity is weakened when a measurement is performed in a time interval of the order or less than $2m$. This could be checked measuring the trajectory of a high energy cosmic ray particle (a proton, for example), freely falling in the gravitational field of the Earth. If the duration of measurement is of the order of $2m$ or less ($m$ being the Earth mass), the trajectory will be less curved. As we already remarked in \cite{HC}, we may now give a reasonable explanation of the fact that the zero point energy does not gravitate: the very fast quantum vacuum fluctuations reduce the strength of gravity so much that its influence is canceled.

\section{Properties of the gravitational fluid}
In order for (2.7) metric (which is not Ricci flat) to be a solution of Einstein's equation $G_{ab} = 8\pi T_{ab}$, with $a,b = t, r, \theta, \phi$, we need a source stress tensor on its r.h.s. The source is an anisotropic fluid with the nonzero components  
  \begin{equation}
T^{r}_{~t} = \frac{m^{2}e^{-\frac{2m}{t}}}{2\pi r^{2}t^{2}},~~~ T^{r}_{~r} = \frac{m}{4\pi rt^{2}}\sqrt{\frac{2m}{r} e^{-\frac{2m}{t}}} = 4 T^{\theta}_{~\theta} =4 T^{\phi}_{~\phi}.    \label{3.1}
 \end{equation}
  Let us take now a congruence of observers with the velocity vector field
  \begin{equation}
  u^{a} = \left(1, - \sqrt{\frac{2m e^{-\frac{2m}{t}}}{r}}, 0, 0\right) ,~~~u^{a}_{~a} = - 1.
 \label{3.2}
 \end{equation}
  The above congruence of observers is geodesic, namely the acceleration $a^{b} = u^{a}\nabla_{a} u^{b} = 0$, to whom the inward radial velocity $ u^{r} = - \sqrt{\frac{2m}{r}e^{-\frac{2m}{t}}}$ is the Newtonian escape velocity. The spacetime (3.1) being nonstatic, the scalar expansion is nonzero
     \begin{equation}
		\Theta \equiv \nabla_{a}u^{a} = -\frac{3}{2r} \sqrt{\frac{2m e^{-\frac{2m}{t}}}{r}}
 \label{3.3}
 \end{equation}
which vanishes when $t \rightarrow 0$ and $r \rightarrow 0$ ( which goes to zero simultaneously with $t$). We also obtain a nonzero shear tensor with the nonzero components $\sigma^{r}_{~r} = -2\sigma^{\theta}_{~\theta} = -2\sigma^{\phi}_{~\phi} = -(2/3)\Theta$ and $\sigma^{r}_{~t} = 2m e^{-\frac{2m}{t}}/r^{2}$.

Consider the general form of an anisotropic fluid with energy flux
  \begin{equation}
  T_{ab} = (p_{t} + \rho) u_{a} u_{b} + p_{t} g_{ab} + (p_{r} - p_{t}) n_{a}n_{b} +  u_{a} q_{b} + u_{b} q_{a},
 \label{3.4}
 \end{equation}
with $\rho(r,t) = T_{ab}u^{a}u^{b}$ the energy density of the fluid, $p_{r}(r,t)$ - the radial pressure, $p_{t}$ - the pressures on the transversal directions $\theta$ and $\phi$, $n^{a}$ is a spacelike vector orthogonal to $u^{a}$, with $n_{a}u^{a} = 0,~n_{a}n^{a} = 1$, $q^{a}$ is the heat flux with $q_{a}u^{a} = 0$ and it is given by the expression $q^{a} = - T^{a}_{~b}u^{b} - \rho u^{a}$, obtained from (3.4). Using now (3.2) and (3.4), one finds that 
  \begin{equation}
 u_{a} = (-1, 0, 0, 0),~~~ n^{a} = \left(0, 1, 0, 0 \right) ,~~~ n_{a} = \left(\sqrt{\frac{2m e^{-\frac{2m}{t}}}{r}}, 1, 0, 0\right).
 \label{3.5}
 \end{equation}
 In spite of the fact that $T^{r}_{~t} \neq 0$, we get from (3.4) a vanishing energy flux $q^{a} = 0$. That is perhaps related to the geodesic character of the congruence (3.2). From (3.4) one further finds that $\rho = 0, p_{r} = T^{r}_{~r} = 4 p_{t}$. Having now the expressions of the energy density and pressures, it is an easy task to see that 
 the weak energy condition (WEC) ($\rho \geq 0,~\rho + p_{r} \geq 0,~\rho + p_{t} \geq 0$),  null energy condition (NEC) ($\rho + p_{r} \geq 0,~\rho + p_{t} \geq 0)$ and strong energy condition (SEC) ($\rho + p_{r} \geq 0,~\rho + p_{t} \geq 0,~\rho + p_{r} + 2p_{t} \geq 0$) are obeyed. However, the dominant energy condition (DEC) ($\rho > |p_{r}|,~ \rho > |p_{t}|$) is not satisfied because $\rho$ is vanishing.

One observes that all components of $T^{a}_{~b}$ vanish when $t \rightarrow \infty$ (or when $t>>2m$) because the metric (2.7) becomes Ricci-flat. We must remind that the limit $r \rightarrow 0$ goes simultaneously with $t \rightarrow 0$ so that $T^{a}_{~b}$ tends to zero at this limit, too. That takes place because of the exponential factor $e^{-\frac{2m}{t}}$ which is present in all expressions, including the scalar curvature $R^{a}_{~a} = -12\pi p_{r}$. Moreover, in the latter case ($t \rightarrow 0$), the geometry (2.7) becomes Minkowskian and the effective mass $m(t)$ goes to zero.

Having now the components of the stress tensor and the basic physical quantities associated to it, our next task is to compute the total energy flow measured by an observer sitting at r = const. \cite{HC3}
 \begin{equation}
 E = \int{T^{a}_{~b}u^{b}n_{a}\sqrt{-\gamma}}dt~d\theta~d\phi,
 \label{3.6}
 \end{equation}
where  $\gamma$ is the determinant of the 3-metric of constant $r$, i.e. $\gamma = -(1 - 2m(t)/r)r^{4} sin^{2}\theta$. With $T^{r}_{~t}$ from (3.1), Eq. (3.6) gives us $E = 0$. The fact that $E = 0$ is not surprising if we remember that P-G observers are in free fall (the acceleration vector $a^{b} = 0$) and the energy flux $q^{a}$ is vanishing.

\section{Conclusions}
The role of the measurement process in gravitational physics is investigated in this paper. 
In the time dependent spacetime we have proposed, the time variable plays the role of the duration of measurement upon some physical system. Very short time intervals w.r.t. the gravitational radius lead to much weaker values of the gravitational field where our system is located. That may direct us to an explanation of the well-known fact that the vacuum energy does not gravitate: very fast quantum fluctuations get rid of the influence of gravity. We also notice that some results from this paper are much more simple than similar quantities obtained in \cite{HC} and all parameters derived are finite throughout.

\end{document}